\documentclass[aps,prd,letterpaper,11pt,preprintnumbers,twoside,tightenlines,nofootinbib,showpacs,preprint]{revtex4-1}
\usepackage{graphicx}
\usepackage[sort&compress]{natbib}
\usepackage{subfigure}
\usepackage{amsmath}
\usepackage{amsfonts}
\usepackage{cancel}
\usepackage{lmodern,dsfont}
\usepackage{amssymb}
\begin{document}
\arraycolsep1.5pt
\newcommand{\Ima}{\textrm{Im}}
\newcommand{\Rea}{\textrm{Re}}
\newcommand{\mev}{\textrm{ MeV}}
\newcommand{\be}{\begin{equation}}
\newcommand{\ee}{\end{equation}}
\newcommand{\ba}{\begin{eqnarray}}
\newcommand{\ea}{\end{eqnarray}}
\newcommand{\gev}{\textrm{ GeV}}
\newcommand{\nn}{{\nonumber}}
\newcommand{\dtres}{d^{\hspace{0.1mm} 3}\hspace{-0.5mm}}
\newcommand{\rts}{ \sqrt s}
\newcommand{\non}{\nonumber \\[2mm]}

\title{The $KD$, $\eta D_s$ interaction in finite volume and the  nature of the $D_{s^* 0}(2317)$ resonance}

\author{A. Mart\'inez Torres$^1$, L. R. Dai$^{1,2}$, C. Koren$^{1,3}$, D. Jido$^1$ and E. Oset$^{1,3}$}
\affiliation{
$^1$Yukawa Institute for Theoretical Physics, Kyoto University,
Kyoto 606-8502, Japan.\\
$^2$Department of Physics, Liaoning Normal University, Dalian, 116029, China.\\
$^3$Departamento de F\'{\i}sica Te\'orica and IFIC, Centro Mixto Universidad de Valencia-CSIC,
Institutos de Investigaci\'on de Paterna, Aptdo. 22085, 46071 Valencia,
Spain
}

\preprint{YITP-11-80}

\date{\today}

 \begin{abstract} 
 An SU(4) extrapolation of the chiral unitary theory in coupled channels done to study the scalar mesons in the charm sector is extended to produce results in finite volume. The theory in the infinite volume produces dynamically the $D_{s^*0}(2317)$ resonance by means of the coupled channels $KD$, $\eta D_s$.  Energy levels in the finite box are evaluated and, assuming that they would correspond to lattice results, the inverse problem of determining the bound states and phase shifts in the infinite volume from the lattice data is addressed.  We observe that it is possible to obtain accurate $KD$ phase shifts and the position of the $D_{s^*0}(2317)$ state, but it requires the explicit consideration of the two coupled channels in the analysis if one goes close to the $\eta D_s$ threshold. We also show that the finite volume spectra look rather different in case the $D_{s^*0}(2317)$ is a composite state of the two mesons, or if it corresponds to a non molecular state with a small overlap with the two meson system. We then show that a careful analysis of the finite volume data can shed some light on the nature of the $D_{s^*0}(2317)$ resonance as a $KD$ molecule or otherwise.

\end{abstract}

\maketitle

\section{Introduction}
\label{Intro}
One of the challenges of lattice QCD calculations is to determine spectra of mesons and baryons. Although problems stem when addressing excited states which have decay channels, steady progress is being done in the field \cite{Nakahara:1999vy,Mathur:2006bs,Basak:2007kj,Bulava:2010yg,Morningstar:2010ae,Foley:2010te,Alford:2000mm,Kunihiro:2003yj,Suganuma:2005ds,Hart:2006ps,Wada:2007cp,Prelovsek:2010gm}. In the case of resonances, the identification of the states obtained in the box with those in the infinite volume is not automatic since scattering states show up as discrete states in the box. The problem is formally solved when the states decay in one channel and the framework of L\"uscher allows to produce phase shifts in the infinite volume, and eventually investigate resonance properties, from the levels in the box \cite{luscher,Luscher:1990ux}. 

Recently, the L\"uscher approach has been generalized to the case of
multi-channel scattering.  This was done in Refs.~\cite{Liu:2005kr} on the
basis of potential scattering theory, while the authors of
Refs.~\cite{Lage:2009zv,akaki} use non-relativistic effective field theory
(EFT) for this purpose. More recently the ideas of Ref.~\cite{akaki} in the meson scalar sector and of Ref.~\cite{Bernard:2008ax}, where the case of the $\Delta$-resonance in $\pi
N$ elastic scattering is addressed, have been followed up in Ref.~\cite{misha}, where a new and practical rederivation of the approach has been done, generalizing it to include fully relativistic effects. Furthermore, new methods of analysis of lattice data are presented for the case of two coupled channels and a thorough investigation of the propagation of errors is done. The results of Ref.~\cite{misha} were done for  $\pi \pi$ and $K \bar{K}$ scattering, addressing the issue of the $f_0(600)$, $f_0(980)$, and $a_0(980)$ resonances. The ``synthetic" lattice data were produced using the chiral unitary approach (where these resonances are dynamically generated) in a finite volume. A follow up of Ref.~\cite{misha} using input from the Juelich model \cite{a}, which requires discretization in momentum space, is done in Ref.~\cite{b} and further work concerning the $\kappa(800)$ scalar resonance is done in Ref.~\cite{Michaelnew}.

   In the present case we face the problem of the scalar mesons in the charmed sector, concretely the strangeness $S=1$, charm $C=1$, where an extremely narrow  resonance, the  $D_{s^*0}(2317)$, appears. This problem has been looked at in the literature studying the interaction of coupled meson-meson channels in this sector and the resonance is found to be dynamically generated, mostly from the interaction of the $KD$, $\eta D_s$ channels \cite{Kolomeitsev:2003ac,Hofmann:2003je,Guo:2006fu,daniel}. An analysis of scalar form factors determined from lattice calculations \cite{juan} finds a large $KD$ scattering length hinting at the existence of a bound $KD$ state which is also identified with the $D_{s^*0}(2317)$. Additional work is done along the lines of Refs.~\cite{Kolomeitsev:2003ac,Hofmann:2003je,Guo:2006fu,daniel} in Ref.~\cite{hanhart2} including the NLO chiral Lagrangian for the interaction of pseudoscalar charmed mesons with Goldstone bosons~\cite{guo1}. The bound state of the $D_{s^*0}(2317)$ is also found and the $KD$ scattering length is evaluated. For other heavy meson-light meson channels, comparison of the scattering length with QCD lattice results is done and good agreement is found.

   In the present work we take the ``phenomenological" potential of Ref.~\cite{daniel} and use the $KD$, $\eta D_s$ channels to evaluate the energy levels in a box using periodic boundary conditions. In a second step, we take these ``synthetic" lattice data and address the inverse problem of extracting from them the $KD$ phase shifts in the infinite volume and the position of the $D_{s^*0}(2317)$, which appears as a bound state of the system. We show that with a few data from the lattice (about ten)  corresponding to two levels of the box with various values of $L$  (the box size), we can reproduce the results in the infinite volume, both for the energy of the bound state and the $KD$ phase shifts, with a good accuracy. For this purpose we use the method with two channels developed in Ref.~\cite{misha}, but generalize the problem to use two-meson loop functions in dimensional regularization rather than in the cut off method used in Ref.~\cite{misha}. We observe that the consideration of the two channels in the analysis is necessary for an accurate reproduction of the data in the infinite volume. Further we show that using several instruments of analysis one can conclude from the synthetic data that the $D_{s^*0}(2317)$ state corresponds mostly to a bound state of $KD$ and is not a genuine state in the sense used by Weinberg \cite{weinberg} to decide that ``the deuteron is not an elementary particle" or by the authors of Ref.~\cite{hanhart} to decide that the ``$a_0(980)$ and $f_0(980)$ are not elementary particles". This means that the analysis of future lattice data in this sector could bring an answer to this question in the case of the $D_{s^*0}(2317)$ state. For this purpose we find out which data are needed and with which precision the energy levels must be evaluated. These results are very opportune since the Tokyo lattice group is addressing this problem at present \cite{ssasaki}.

\section{Formalism}
\subsection{The chiral unitary approach in infinite volume}

 In the chiral unitary approach the scattering matrix in coupled channels is given by

\be
T=[1-VG]^{-1}V
\label{bse}
\ee
where $V$ is the matrix for the transition potentials between the channels and $G$, a diagonal matrix, is in our case the loop function of two meson propagators, which is defined as 

\begin{equation}
G=i\int \frac{d^4 q}{(2\pi)^4}\frac{1}{q^2-m^2+i\epsilon}\frac{1}{(P-q)^2-M^2+i\epsilon}\,
\label{loop}
\end{equation}
where $m$ and $M$ are the masses of the two mesons and $P$ the fourmomentum of the external meson-meson system.

In dimensional regularization this integral is evaluated and gives for meson-meson systems \cite{ollerulf,bennhold}

\begin{eqnarray}
\label{eq:g-function}
 \mbox G_i(s, m_i, M_i) &=&  \frac{1}{(4 \pi)^2}
  \left\{
        a_i(\mu) + \log \frac{m_i^2}{\mu^2} +
        \frac{M_i^2 - m_i^2 + s}{2s} \log \frac{M_i^2}{m_i^2}
  \right.
  \\
     &+& \frac{Q_i(\rts)}{\rts}
    \left[
         \log \left(  s-(M_i^2-m_i^2) + 2 \rts Q_i(\rts) \right)
      +  \log \left(  s+(M_i^2-m_i^2) + 2 \rts Q_i(\rts) \right)
    \right.
  \nonumber \\
  &- &
  \Biggl.
    \left.
    \log \left( -s+(M_i^2-m_i^2) + 2 \rts Q_i(\rts) \right)
      - \log \left( -s-(M_i^2-m_i^2) + 2 \rts Q_i(\rts) \right)
    \right]
  \Biggr\},
  \nonumber
\end{eqnarray}
where $s=E^2$, with $E$ the energy of the system in the center of mass frame, $Q_i$ the on shell momentum of the particles in the channel, $\mu$  a regularization scale and $a_i(\mu)$ a subtraction constant (note that there is only one independent parameter, because a change in $\mu$ can be absorbed into $a_i$).

In other works one uses regularization with a cut off in three momentum once the $q^0$ integration is analytically performed \cite{npa} and one gets 

\ba
G_j&=&\int\limits^{|\vec q|<q_{\rm max}}
\frac{d^3\vec q}{(2\pi)^3}\frac{1}{2\omega_j(\vec q)\,\omega^\prime_j(\vec q)}
\frac{\omega_j(\vec q)+\omega^\prime_j(\vec q)}
{E^2-(\omega_j(\vec q)+\omega^\prime_j(\vec q))^2+i\epsilon},
\non 
\omega_{j}(\vec q)&=&\sqrt{m_{j}^2+\vec q^{\,\,2}},\quad \omega^\prime_{j}(\vec q)=\sqrt{M_{j}^2+\vec q^{\,\,2}}\,
\label{prop_cont}
\ea
In Ref.~\cite{ollerulf} the equivalence of the two methods was established.

We consider s-wave coupled-channels of $KD$ and $\eta D_s$ with $I=0$ for the description of the $D_{s^*0}(2317)$ as a dynamically generated state. The third channel considered in Ref.~\cite{daniel}, $\eta_c D_s$, plays a negligible role and is not considered here. The pseudoscalar mesons $K$ and $D$ are represented as the doublet of isospin 

\be
\Bigg(\begin{array}{c}
K^+\\ 
K^0
\end{array}\Bigg)\,\,\, ,
\,\,\,
\Bigg(\begin{array}{c}
D^+\\ 
-D^0
\end{array}\Bigg)
\ee

The isospin $I=0$ $KD$ state is given by
\be
|KD, I=0> = -\frac{1}{\sqrt 2} |K^+ D^0 + K^0 D^+>
\ee

The state of $\eta D_s$ is already a state of $I=0$. By naming the channels, 1 for $KD$ and 2 for $\eta D_s$, the transition potentials are given by \cite{daniel}
\begin{align}
V_{11}&= - \frac{1}{3f_\pi f_D}[\gamma(t-u) + s-u +m_D^2+m_K^2]
\non
V_{22}&= - \frac{1}{9f_\pi f_D}[ \gamma(-s+2t-u) +2m_D^2+6m_K^2-4m_\pi^2]
\non
V_{12}&= \frac{1}{6\sqrt3 f_\pi f_D}[\gamma (u-t)-(3+\gamma)(s-u) -m_D^2-3m_K^2+2m_\pi^2]
\label{potential}
\end{align}
where 
\be
\gamma= \Bigg (\frac{m_L}{m_H}\Bigg)^2=\Bigg(\frac{m_\rho}{m_{D^*}}\Bigg)^2
\ee

We study only the s-wave interaction and, hence, we must project $t$ and $u$ over s-wave, which is given by 
\be
\bar{u}=m_1^2+m_4^2- \frac{1}{2s}(s+m_1^2-m_2^2)(s+m_4^2-m_3^2)
\ee
where $m_i$ are the masses for the process $1,2 \to 3,4$.
The variable $t$ projected over s-wave, $\bar{t}$, can be obtained from the condition

\be
s+\bar{u}+\bar{t}= m_1^2+m_2^2+m_3^2+m_4^2
\ee

The use of Eq.~(\ref{bse}) with the two channels that we consider leads to a dynamically generated state using the dimensionally regularized $G^D$ function with $\mu=1500$ MeV and the subtraction constant $a=-1.26$ at the energy of 2317 MeV, which we associate to the $D_{s^*0}(2317)$ resonance.

\subsection{The chiral unitary approach in a finite box}
When one wants to obtain the energy levels in the box, one replaces the $G$ function by a $\tilde G$ in Eq.~(\ref{bse}), where instead of integrating over the energy states of the infinite volume, with $q=|\,\vec{q}\,|$ being a continuous variable, as in Eq.~(\ref{prop_cont}), one sums over the discrete momenta allowed in a finite box of side $L$ with periodic boundary conditions. We then have $\tilde G={\rm diag}\,(\tilde G_1,\tilde G_2)$, where 
\ba
\tilde G_{j}&=&\frac{1}{L^3}\sum_{\vec q}^{|\vec q|<q_{\rm max}}
\frac{1}{2\omega_j(\vec q)\,\omega^\prime_j(\vec q)}\,\,
\frac{\omega_j(\vec q)+\omega^\prime_j(\vec q)}
{E^2-(\omega_j(\vec q)+\omega^\prime_j(\vec q))^2},
\non 
\vec q&=&\frac{2\pi}{L}\,\vec n,
\quad\vec n\in \mathds{Z}^3 \,
\label{tildeg}
\ea

 This is the procedure followed in Ref.~\cite{misha}. 

Since the work of Ref.~\cite{daniel} was done using $G$ functions in dimensional regularization, we are going to adapt the finite volume formalism to the use of the dimensionally regularized $G$ functions in the infinite volume. For this purpose let us write the $G$ function of Eq.~(\ref{eq:g-function}) as 

\begin{align}
G^D(E)=b+\int\limits_{|\vec{q}|<\mu}\frac{d^3q}{(2\pi)^3}I(q)+\lim_{q_{\rm max}\to \infty}\int\limits_{\mu}^{q_{\rm max}}\frac{d^3q}{(2\pi)^3} I(q)-\lim_{q_{\rm max}\to \infty}\int\limits_{\mu}^{q_{\rm max}}\frac{d^3q}{(2\pi)^3} I^a(q)
\label{gdsep}
\end{align}
where $I(q)$ is the integrand of Eq.~(\ref{prop_cont})

\ba
I(q)&=&\frac{1}{2\omega_j(\vec q)\,\omega^\prime_j(\vec q)}
\frac{\omega_j(\vec q)+\omega^\prime_j(\vec q)}
{E^2-(\omega_j(\vec q)+\omega^\prime_j(\vec q))^2+i\epsilon},
\label{prop_contado}
\ea
with $\omega_j$, $\omega^\prime_j$ defined in Eq.~(\ref{prop_cont}) and $I^a(q)$ the asymptotic expression of $I(q)$ when $q$ goes to infinity.

\begin{align}
I^a(q)=\frac{1}{2q^2}\frac{2q}{-(2q)^2+i\epsilon}=\frac{1}{-4q^3}
\label{asymptotic}
\end{align}

The constant $b$ in Eq.~(\ref{gdsep}) is a remnant subtraction constant inherent to the regularization procedure. Thus we can write

\begin{align}
G^D(E)&=b+\int\limits_{|\vec{q}|<\mu}\frac{d^3q}{(2\pi)^3}I(q)+\lim_{q_{\rm max}\to \infty}\Bigg[\int\limits_{\mu}^{q_{\rm max}}\frac{d^3q}{(2\pi)^3} I(q)+\frac{1}{8\pi^2}\ln\Bigg(\frac{q_{\rm max}}{\mu}\Bigg)\Bigg]\nonumber\\
&=b+\lim_{q_{\rm max}\to \infty}\Bigg[\int\limits_{q<q_{\rm max}}\frac{d^3q}{(2\pi)^3} I(q)+\frac{1}{8\pi^2}\ln\Bigg(\frac{q_{\rm max}}{\mu}\Bigg)\Bigg]
\label{gdexplicit}
\end{align}

When we study the problem of the energy levels in the box we will substitute $G^D$ by $\tilde G$, replacing the integral by the discrete sum of Eq.~(\ref{tildeg}) up to the same $q_{max}$, and since from Eq.~(\ref{gdexplicit}) we have in the limit of $q_{max} \to \infty$

\begin{align}
b+\frac{1}{8\pi^2}\ln\Bigg(\frac{q_{\rm max}}{\mu}\Bigg)=G^D(E)-\int\limits_{q<q_{\rm max}}\frac{d^3q}{(2\pi)^3} I(q)
\label{substitute}
\end{align}
we can write

\begin{align}
\tilde{G}(E)=G^D(E)+\lim_{q_{\rm max}\to \infty}\Bigg[\frac{1}{L^3}\sum_{q_i}^{q_{\rm max}}I(q_i)-\int\limits_{q<q_{\rm max}}\frac{d^3q}{(2\pi)^3} I(q)\Bigg]
\label{gtdim}
\end{align}

The three dimensional sum in Eq.~(\ref{gtdim}) can be reduced to one dimension considering the multiplicities of the cases having the same $\vec{n}^{\,2}$  \cite{b,c}. The integral in Eq.~(\ref{gtdim}) has an analytical form as shown in the appendix of  Ref.~\cite{d}  (see erratum).

\begin{figure}
\includegraphics[width=0.65\textwidth]{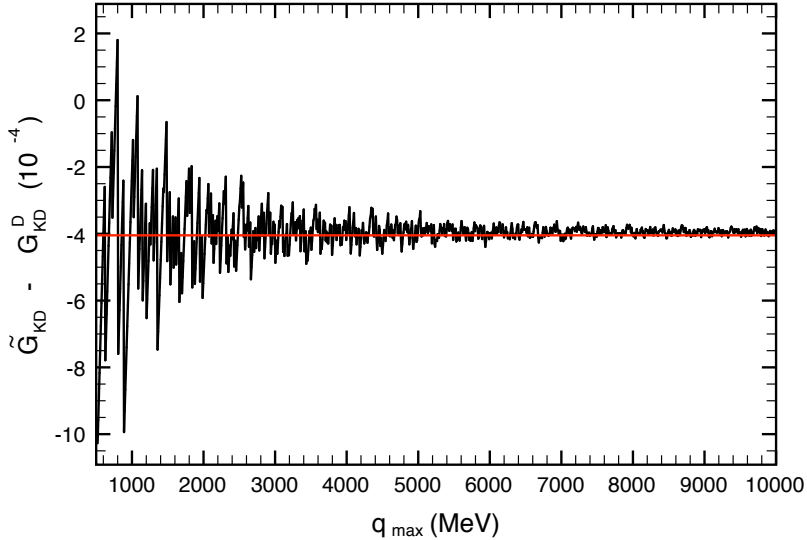}
\caption{Real part of the last two terms of the right hand side of Eq.~(\ref{gtdim}) for the $KD$ channel. The solid line indicates the average that we take between 1200 MeV and 2000 MeV for $q_{max}$. The results correspond to a value of $L= 2.4~m_\pi^{-1}$ and E= 2317~MeV.}
\label{fig:oscilation}
\end{figure}

When calculating the limit of $q_{max}$ going to infinity in Eq.~(\ref{gtdim}) we obtain oscillations which gradually vanish as $q_{max}$ goes to infinity. Yet, it is unnecessary to go to large values of $q_{max}$, and performing an average for different 
$q_{max}$ values between 1200 MeV and 2000 MeV one obtains a perfect convergence, as one can see in Fig. \ref{fig:oscilation}.  Note that the imaginary part of $G^D$ and of the integral in Eq.~(\ref{gtdim}) are identical and they cancel in the construction of $\tilde G$, which is a real function.  

The eigenenergies of the box correspond to energies  that produce poles in the $T$ matrix.  Thus we search for these energies by looking for zeros of the determinant of $1-V\tilde G$

\be
\label{eq:det}
\det(1-V\tilde G)=1-V_{11}\tilde G_1-V_{22}\tilde G_2
+(V_{11}V_{22}-V_{12}^2)\tilde G_1\tilde G_2=0\, .
\ee

\begin{figure}
\includegraphics[width=0.65\textwidth]{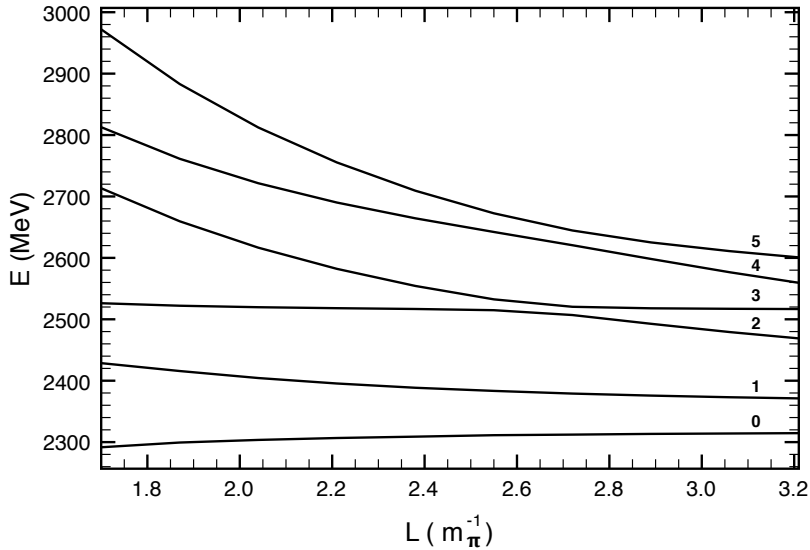}
\caption{Energy levels as functions of the cubic box size $L$, derived 
from the coupled channels unitary approach of Ref.~\cite{daniel} and using 
$\,\tilde G$ from Eq.~(\ref{gtdim}).}
\label{fig:levelstwo}
\end{figure}

In Fig. \ref{fig:levelstwo} we show the energy levels obtained for the box for different values of $L$. We show there the first 5 levels. We observe a smooth behavior of the levels as a function of $L$. The lowest level converges for large $L$ to the value of the energy of the bound state of the infinite volume case. This is a standard feature for the ground level bound state in the lattice, where the energy obtained is about the same as in the infinite volume for sufficiently large values of $L$.  The first valuable information from our study is how large should $L$ be to provide an accurate value of the energy. We find that with values of $L=3~ m_\pi^{-1}$ there is already a good convergence to the value in the infinite volume. Yet, we would like to get more information from the lattice data, for instance scattering phase shifts of $KD$, and eventually $\eta D_s$, but we will restrict ourselves to just the first channel. We could also ask ourselves about the nature of the bound state found.  Of course, in the present problem where the $D_{s^*0}(2317)$ appears dynamically generated we should get as an answer that it is indeed a bound state of $KD$ with some admixture of $\eta D_s$, but if the lattice data were different than the levels obtained by us, the possibility exist that the answer would be different.  We address these problems in section V.
\section{The inverse problem of getting phase shifts from lattice data}
\subsection{Results with two channels}

In this section we face the problem of getting bound states and phase shifts in the infinite volume from the energy levels obtained in the box using the two channel approach of Ref.~\cite{daniel}, which we would consider as ``synthetic" lattice data. To accomplish this we need more information than just the lowest level, but we shall see that the first two levels shown in Fig.~\ref{fig:levelstwo} already provide the necessary information to reproduce the problem in the infinite volume.

 In Ref.~\cite{misha} several methods were suggested to solve the inverse problem, but we borrow here the one based on a fit to the data in terms of a potential suggested by the work of Ref.~\cite{daniel} or  Ref.~\cite{Kolomeitsev:2003ac,Hofmann:2003je,Guo:2006fu}. As we can see in Eq.~(\ref{potential}), the potentials have a large constant part, some terms proportional to $s$ and some terms inversely proportional to $s$. It is very easy to see that if one chooses a region of energies around a certain value of $s$, $s_0$, the inverse function of $s$ can be expanded as a function of $s-s_0$ to a good approximation. Choosing $s_0= (m_K +M_D)^2$ then the ansatz of the following equation 

\be
V_{ij}=a_{ij}+b_{ij}(s-(m_K +M_D)^2)\, 
\label{fitv}
\ee
is a very accurate assumption.  Comparing Eq.~(\ref{fitv}) with Eq.~(\ref{potential}), used in the chiral unitary approach of Ref.~\cite{daniel}, we find 

\begin{align}
a_{11}&=-155.101, \quad b_{11}=-3.732\times 10^{-5} \textrm{MeV}^{-2}\nonumber\\
a_{12}&=a_{21}=-90.756, \quad b_{12}=b_{21}=-3.361\times 10^{-5} \textrm{MeV}^{-2}\label{para}\\
a_{22}&=-52.356, \quad b_{22}=9.395\times 10^{-7} \textrm{MeV}^{-2}\nonumber
\end{align}

Certainly, this is not the unique option to face the inverse problem. In Ref.~\cite{misha} it was already discussed that one could use a different parametrization and the consequences of it were discussed there. This problem has been studied in more detail in Ref.~\cite{Michaelnew} where a systematic study of results obtained with different options of the potential is done. The conclusion of that work is that the results obtained with the simple form of the potential, Eq.~(\ref{fitv}), are fairly good, but the freedom to use other potentials reverts into a somewhat larger uncertainty in the parameters of the resonances searched.

\begin{figure}
\includegraphics[width=0.65\textwidth]{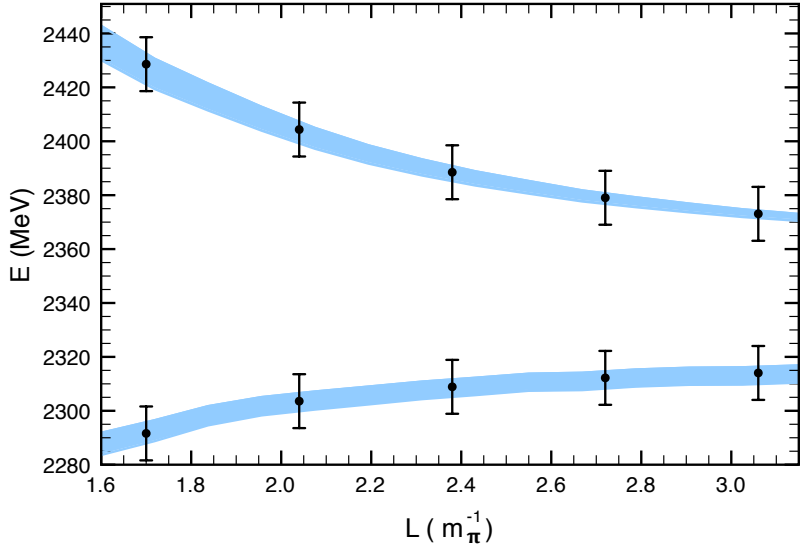}
\caption{Energy levels as functions of the cubic box size $L$, reconstructed from fits to the ``data" of Fig. \ref{fig:levelstwo} using the potential of Eq.~(\ref{fitv}). The band corresponds to different choices of parameters within errors.}
\label{fig:levelstworecons}
\end{figure}

We assume that the lattice studies provide us with ten eigenenergies corresponding to the first two levels of Fig.~\ref{fig:levelstwo} for different values of $L$ between $1.7~m_{\pi}^{-1}$ and $3.3~m_{\pi}^{-1}$. We also assume that the levels are provided with an error of $\pm 10$  MeV, something achievable in present QCD lattice calculations. 
We make a best fit to the data assuming a potential as in Eq.~(\ref{fitv}). We look for the minimum $\chi^2$  and obtain a set of parameters for $a_{ij}, b_{ij}$. Then we generate random sets of the parameters close to those of the minimum, such that $\chi ^2$ is only increased below $\chi_{min}^2+ 1$, a criteria that provides a band of fair statistical errors \cite{misha,Ben}. With these values we generate the spectrum of Fig. \ref{fig:levelstworecons} by searching for the zeros of the determinant of $1-V\tilde G$. It might look like this is a tautology, since by using the same potential as in the original infinite volume problem and the same subtraction constant in $\tilde G$ we should obtain for the parameters the same results as in the original potential (see Eq.~(\ref{para})) and $\chi^2_{min}=0$. And this is indeed the case. However, when the lattice data are provided to us we do not know which implicit regularization subtraction constant the lattice data would support. The inverse method only has a real value if the results that one obtains are independent of this subtraction constant. In one channel the improved L\"uscher approach of Ref.~\cite{misha} shows clearly that the results of physical quantities, like the phase shifts, are independent of the cut off chosen, for values of $q_{max}$ relatively larger than the on shell momenta of the particles ($q_{max} > 1.5$ GeV in general), and has a defined limit for $q_{max}$ going to infinity. In the two coupled channels it was also found in Ref.~\cite{misha} that the results were basically cut off independent, because a change in the regularization scheme should be absorbed by the fitted coefficients of the potential, Eq.~(\ref{fitv}), \cite{e}. This is related to the basic feature of the renormalization group, with a trade off between the scale and the potential \cite{friman}. This finding is what renders very valuable the approach of Ref.~\cite{misha} to solve the inverse problem.  The arbitrariness in the choice of the cut off in Ref.~\cite{misha} translates here in the arbitrariness in the choice of the subtraction constant $a$ in the analysis of the inverse problem, and we also find that the results are independent of this choice. This means that we choose an arbitrary subtraction constant in $\tilde G$ in the analysis of the lattice data, and the same one in $G^D$ to reconstruct the phase shifts in the infinite volume. Like in the cut off method, we also take a range for the subtraction constant of $a \in [-1,-2]$, considered of natural size in Refs. \cite{ollerulf} and \cite{daniel}.

\begin{figure}
\includegraphics[width=0.65\textwidth]{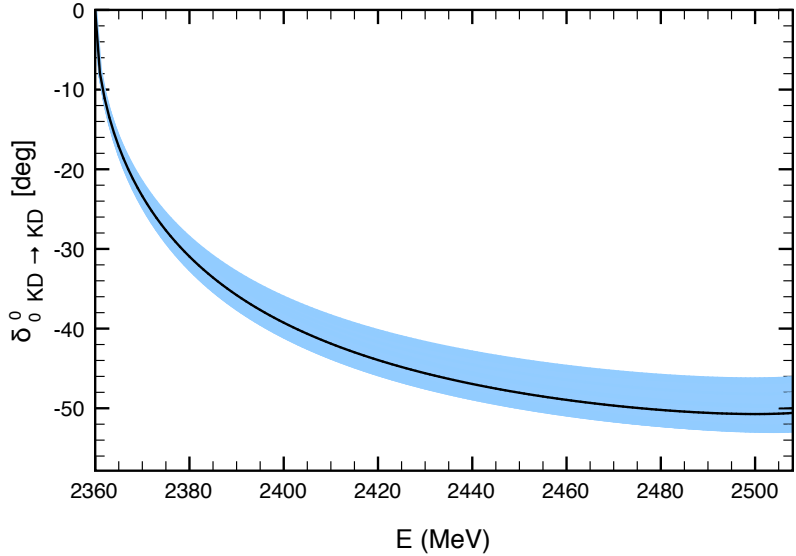}
\caption{Phase shifts for $KD$ scattering derived 
from the coupled channels unitary approach of Ref.~\cite{daniel} (solid line). The band corresponds to the results obtained from the fits to the "data" of Fig. \ref{fig:levelstwo} using the potential of Eq.~(\ref{fitv}) with two channels.}
\label{fig:phaseshiftstwo}
\end{figure}

As we mentioned above, in Fig. \ref{fig:levelstworecons} we show now the results of the levels reconstructed from the best fits to the data, with a band corresponding to the random choices of parameters satisfying the condition that $\chi ^2 < \chi_{min}^2+ 1$. More interesting is to construct from these levels the $KD$ phase shifts in the infinite volume. This can be seen in Fig. \ref{fig:phaseshiftstwo}.  The phase shifts are evaluated from the $T_{11}$ matrix element obtained from the Bethe-Salpeter equation, Eq.~(\ref{bse}), using the potentials obtained in the fit and the continuum $G^D$ function. The normalization that we use is such that in one channel \cite{npa} 

\be
T(E)=\frac{-8\pi E}{p\cot \delta(p)-i\,p}\, ,
\ee
from where we determine the phase shifts in the infinite volume problem.

In Fig. \ref{fig:phaseshiftstwo} we can see the $KD$ s-wave phase shifts for $I=0$, $\delta^0_0$, reconstructed with our procedure for the different values of the parameters generated. As we can see, the agreement with the exact results is quite good, and we see how the errors in the determination of the lattice levels have propagated in the determination of the phase shifts. 

\subsection{Results with one channel analysis}
In the present problem the $\eta D_s$ channel is far away from the $KD$ threshold and the channel $KD$ is more important than the $\eta D_s$ for the bound state and for low energies of the $KD$ scattering \cite{daniel}. We may wonder whether a fit to the lattice data would be possible with only one channel. It is well known that the effects of far away channels which are not too relevant in a problem can be incorporated with modifications in the potential or $G$ function of the main channel. This is sometimes done explicitly, like in Ref.~\cite{hyodo}, where an effective $\bar K N $ interaction is constructed that incorporates the effects of the $\pi \Sigma$ channel, even if this latter channel is relevant. The reason for this is that if a second channel is far away from the region of energies studied, its effect in this region is fairly energy independent (changes in $\Delta E$ of order $\Delta E/ (E_2-E)$, where $E_2$ is the threshold of this second channel and $E$ the energy under investigation). Then the effect of this channel is easily accounted for by a slight modification in the subtraction constant of the first channel. Yet, as we approach the threshold of the second channel this trade off does not hold and the explicit need of that channel becomes manifest.

   For the one channel problem, as shown in Ref.~\cite{misha}, the $T$-matrix in the infinite volume can be obtained for the energies which are eigenvalues of the box by 
   
   \be
T(E)=\left(V^{-1}(E)-G^D(E)\right)^{-1}= \left(\tilde G(E)-G^D(E)\right)^{-1} \ . 
\label{extracted_1_channel}
\ee

One can equally use the same procedure as done for two channels eliminating the $V_{12}$ and $V_{22}$ parts of the potential. The results with both methods are basically identical.  By using the same subtraction constant as used to generate the spectra we obtain $a_{11}=-191.954$ and $b_{11}=-9.868\times 10^{-5}$ $\textrm{MeV}^{-2}$ for the best fit, similar to the value in the two channel case. Note, however, that as shown in Eq.~(\ref{tonediff}) the results for $T$ are independent of this regularization constant.

It is interesting to remark that given the structure of $\tilde G$ in dimensional regularization, Eq.~(\ref{gtdim}), in the difference of Eq.~(\ref{extracted_1_channel}) the function $G^D$ cancels identically and one finds

\begin{align}
T(E)^{-1}=\lim_{q_{max}\to \infty}\Bigg[\frac{1}{L^3}\sum_{q_i}^{q_{max}}I(q_i)-\int_{q<q_{max}}\frac{d^3q}{(2\pi)^3} I(q)\Bigg]
\label{tonediff}
\end{align}

This result is the same as the one obtained in Ref.~\cite{misha} starting with cut off regularization, and, as proved in Ref.~\cite{misha} is nothing else than L\"uscher formula, except that Eq.~(\ref{tonediff}) keeps all the terms of the relativistic two body propagator. Indeed L\"usher results have an implicit approximation since some terms of the real part of the two body propagator, which are exponentially suppressed in the physical region, are omitted (see section 2.2 and Appendix A of Ref.~\cite{misha}).

\begin{figure}
\includegraphics[width=0.65\textwidth]{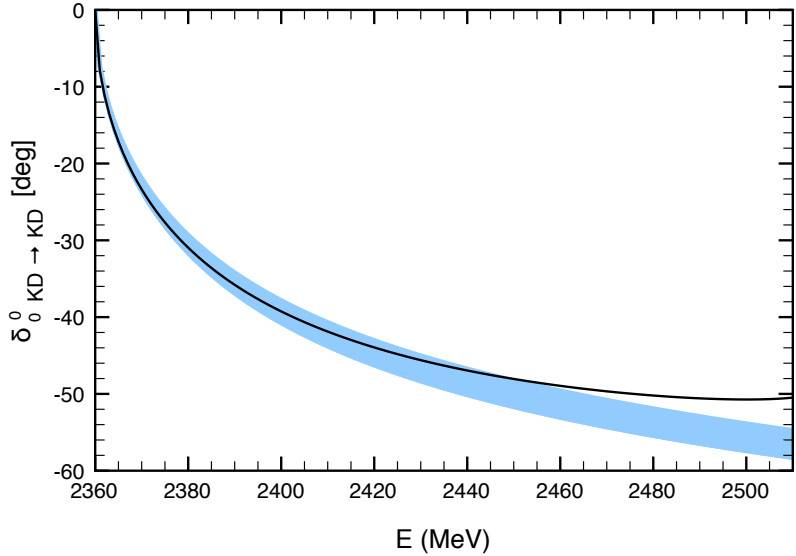}
\caption{Phase shifts for KD scattering derived 
from the coupled channels unitary approach of Ref.~\cite{daniel} (solid line). The band corresponds to using the fits to the ``data" of Fig. \ref{fig:levelstwo} using the potential of Eq.~(\ref{fitv}) with only the  $KD$ channel.}
\label{fig:shiftsonechannel}
\end{figure}
 
 In Fig. \ref{fig:shiftsonechannel} we can see the results for the phase shifts obtained with one channel. We observe that for low energies the results are very similar to those obtained with two channels. As the energy increases, the results with two channels approach better the exact results. 
 
\section{The bound state and its nature}

For the energy of the bound state we obtain also good results with the one and two channel methods. The numerical results are
  $E= 2317 \pm 5 $ MeV in both cases.
  
 It would be interesting to see if from the lattice data we could say something about the nature of the $D_{s^*0}(2317)$ state. Obviously with the ``synthetic" lattice data which we have produced the answer is trivial since the state was dynamically generated in our approach. But if the real lattice results were different than those obtained here the answer might be different.

 \subsection{Dynamically generated states}
 
 The problem has been solved for bound states close to threshold using the method of Weinberg, based on the knowledge of the scattering length and effective range \cite{weinberg}, and also for resonances not far from a threshold \cite{hanhart}. Using this method in Ref.~\cite{misha} it was found that the deuteron was a composite state of neutron and proton. The method, updated to the present problem has also been used in Ref.~\cite{guo2}, where claims are made that knowing the dependence of the $KD$ scattering length on the $K$ mass can help to learn about the nature of the $D_{s^* 0}(2317)$.

We follow here a different approach by going to the root of the derivation of the results of Ref.~\cite{weinberg}.  These results stem from a sum rule that comes from the normalization to unity of the wave function of the bound state.  A modern formulation of this sum rule can be seen in  Ref.~\cite{arriola} and states that
 
 \be
\sum_i g_i^2 \left.\frac{dG_{ii}}{dE}\right|_{E=E_\alpha}=-1
\ee
where $g_i^2$ are the residues of the $T_{ii}$ scattering matrices (coupling squared)  at the pole  of the bound state ($\alpha$) and  $G_{ii}$ are the propagator of the two particles of the corresponding channels  (the loop function $G^D$ that we use here). However, the normalization of the couplings and the $G$ functions in Ref.~\cite{arriola} correspond to a nonrelativistic Quantum mechanics formalism, while here we use a different normalization in the field theoretical approach (the correspondence can be found in Ref.~\cite{arriola}). It is easy to generalize this result for the field theoretical case as
 \be
\sum_i g_i^2 \left.\frac{dG_{ii}}{ds}\right|_{E=E_\alpha}=-1\label{eq76}
\ee
where now the residues are defined as
\be
g_i^2= \lim_{ s \to s_R} (s-s_R)T_{ii}
\label{residuenew}
\ee
with $s_R$ the energy squared of the bound state. A different derivation of this equation is done in Ref.~\cite{jidohyodo} using the Ward Takahashi identity. For the case of one channel this is easy to see since 
\be
T=\frac{1}{V^{-1}-G}
\label{tonechannel}
\ee
and if $V$ is energy independent it follows immediately that 

\be
g^2= -\frac{1}{d G/ d s}
\ee

One can see from the derivation in Ref.~\cite{arriola} that each term in Eq.~(\ref{eq76}) accounts (with reversed sign) for the probability of the bound state to be a bound state of the pair of particles of the channel considered. In the case that we had the coupling of the bound state to another hypothetical elementary particle outside the space of pair of particles considered, there would be an extra term in the sum of Eq.~(\ref{eq76}), -$Z$, where $Z$ would account for the overlap of the bound state with this hypothetical genuine particle. The diversion of the sum of Eq.~(\ref{eq76}) with respect to -1 would indicate the amount of the bound state which cannot be considered a bound state of the two particles. Actually, for the compositeness of a bound state in a single channel, the relation
\begin{equation}
 - g^2 \frac{dG}{ds} = 1 - Z
\end{equation}
was derived with a field theoretical argument~\cite{two}. Here $Z$ is the field renormalization constant for the genuine state introduced to the theory by hand.

In our formulation, there is a caveat, since in the derivation we have assumed that $V$ is energy independent, while we are now producing potentials with a moderate energy dependence. We have checked that if we use a potential independent of the energy with the two channels, the theorem of Eq.~(\ref{eq76}) holds exactly in our calculation. Yet, with the energy dependence of Eq.~(\ref{fitv}) the sum rule provides a value of -0.8. We should not look at this as an indication that we have the coupling to a genuine state. We should first see what comes out from the consideration of the energy dependence which we know explicitly in our problem. Taking into account the energy dependence of Eq.~(\ref{fitv}) we find that approximately one should have now (see also Ref.~\cite{jidohyodo})

\begin{align}
\sum_{i} g^2_i\Bigg\{\frac{b_{ii}}{[a_{ii}+b_{ii}(s-(m_D+m_K)^2)]^2}+\frac{d G_i}{d s}\Bigg\}=-1
\label{sumrulenergy}
\end{align}

The former equation is fulfilled up to the level of 3\%, hence its application could tell us, if the diversion from -1 of the sum rule is appreciable, that there would be a substantial coupling to a genuine state in the present work. Due to the construction of the state in the present problem, this is obviously not the case, as one can see by the proximity to -1 of the sum in Eq.~(\ref{sumrulenergy}). Furthermore, we observe that about 90 \% of the sum rule comes from the $KD$ state, indicating that we have largely a bound $KD$ channel in our approach. The successful analysis of one channel would obviously indicate that the state is 100\% a bound state of $KD$. This shows the difficulty that we might have quantizing with high precision the amount of $KD$ in the bound state of $D_{s^*0}(2317)$ from the analysis of the lattice data, yet, the use of an extra level of the box would allow us to be more precise. 

\subsection{The case of a genuine state}
On the other hand, there is the possibility that the data might be such that one could find a gross deviation from having $KD$ as the large component of the  $D_{s^*0}(2317)$ wave function, reflecting the fact that the $D_{s^*0}(2317)$ would be a genuine state, not generated by the $KD$ interaction.  For this purpose we have made a test introducing a different potential where a CDD pole (Castillejo, Dalitz, Dyson) \cite{leonardo} is introduced by hand.  The potential in just one channel would now be

\begin{align}
V=V_M+\frac{g^2_{CDD}}{s-s_{CDD}}
\label{cdd}
\end{align}
where $V_M$ is assumed to be energy independent and $g^2_{CDD}$, $s_{CDD}$ are the parameters of the CDD pole.

Using Eqs.~(\ref{residuenew}) and (\ref{tonechannel}) and taking into account that for the bound state we have $V^{-1}=G$, we find now (see also Ref.~\cite{two}) 

\begin{align}
-g^2\frac{d G}{d s}&=\frac{1}{-\dfrac{d V^{-1}}{d s}\Big(\dfrac{dG}{ds}\Big)^{-1}+1}=1-Z\nonumber\\
&=\frac{1}{-\dfrac{g^2_{CDD}G^2}{(s-s_{CDD})^2\dfrac{dG}{ds}}+1}
\label{sumrulecdd}
\end{align}
which indeed shows that $Z$ is a quantity between 0 and 1 since  $\dfrac{dG}{ds} < 0$ below the $KD$ threshold.

\begin{figure}
\includegraphics[width=0.65\textwidth]{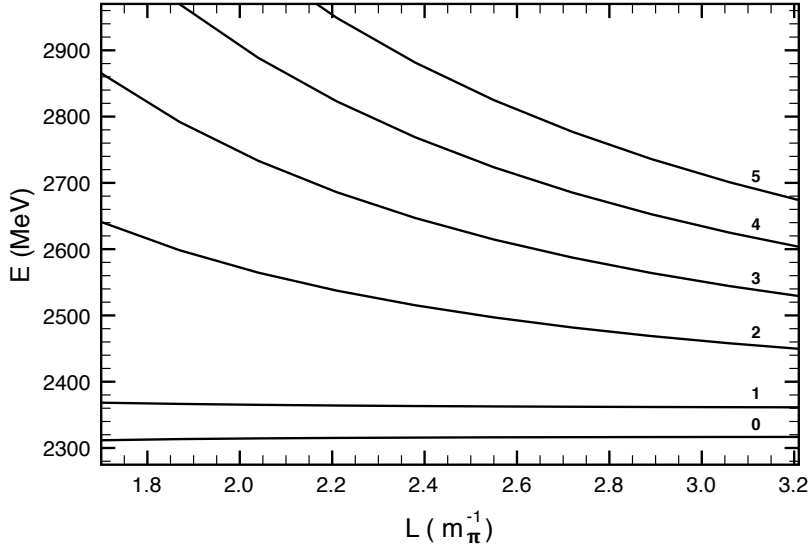}
\caption{Energy levels in the box constructed from the potential of Eq.~(\ref{cdd}) that contains a CDD pole.}
\label{fig:levelscdd}
\end{figure}

We have chosen a potential of the type of Eq.~(\ref{cdd}) with $V_M$ of the order of ten times smaller that the potential used for $V_{11}$, $\sqrt {s_{CDD}}$ corresponding to a 20 MeV above the mass of the $D_{s^*0}(2317)$  and $g_{CDD}=3787.34$ MeV such as to reproduce the bound state at the mass of the $D_{s^*0}(2317)$. We find that at the pole of this state, Eq.~(\ref{sumrulecdd}) provides $Z\sim 0.9$, thus showing that the introduction of a CDD pole as in Eq.~(\ref{cdd}) is a good method to account for a genuine state beyond the space of the two interacting particles.

The levels in the box with this potential are shown in Fig.~\ref{fig:levelscdd}. As we can see, the levels show quite a different behaviour than those found with the potential of the coupled channel unitary approach \cite{daniel} shown in Fig. \ref{fig:levelstwo}. It is clear that the determination of the levels with lattice calculations can easily differentiate between these two scenarios.

\begin{figure}
\includegraphics[width=0.65\textwidth]{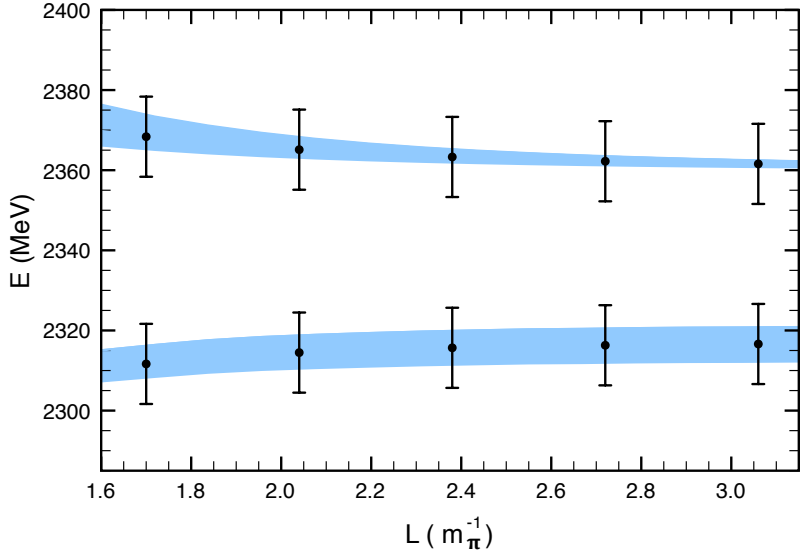}
\caption{Fit to the first two energy levels in the box of Fig. \ref{fig:levelscdd} constructed from the potential of Eq.~(\ref{cdd}) that contains a CDD pole.}
\label{fig:cddfitcdd}
\end{figure}

\begin{figure}
\includegraphics[width=0.65\textwidth]{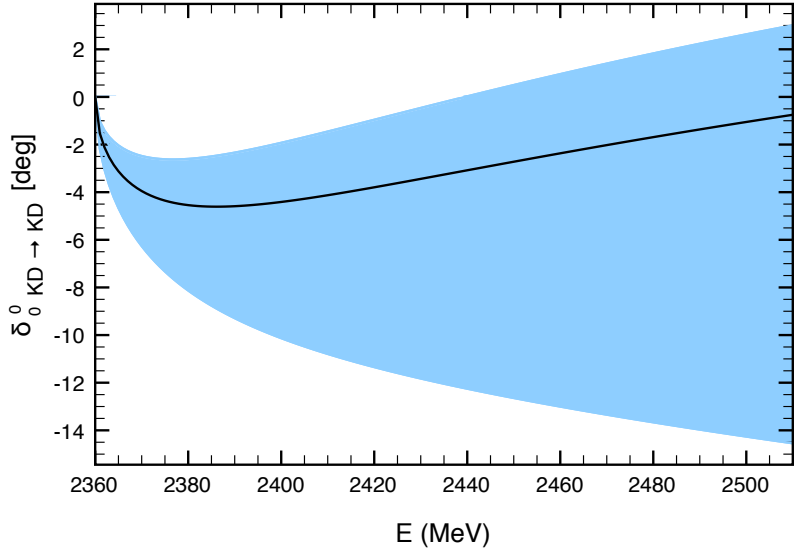}
\caption{Phase shifts for $KD$ from the fits of Fig. \ref{fig:cddfitcdd} using the potential of Eq.~(\ref{cdd}) that contains a CDD pole. The solid line corresponds to the results obtained with the CDD potential of reference, and the band corresponds to the fits to the first two levels with a potential of the same type considered in Fig \ref{fig:cddfitcdd}.}
\label{fig:shiftscddfit}
\end{figure}

In Fig. \ref{fig:cddfitcdd} we show the results with a fit to the first two levels of Fig. \ref{fig:levelscdd} obtained with the CDD pole structure for the potential. We have taken also ten points over the curves and assumed $\pm 10$ MeV errors. With the values of the parameters that fulfill $\chi^2 <\chi_{min}^2 + 1$ we show in Fig. \ref{fig:shiftscddfit} the phase shifts that one would obtain.  We observe that the phase shifts are quite different from those obtained in the two channels with the standard potential.  

At this point we would like to comment that the phase shifts obtained with the standard potential, Eq.~(\ref{fitv}), or with the CDD potential, Eq.~(\ref{cdd}), are negative close to the $KD$ threshold. According to Ref.~\cite{newsasaki} (see also Ref.~\cite{sasakifriend}) this is related to the presence of the bound state and Levinson's theorem which states that the difference of the phase shift between zero energy and infinity, for one channel with ordinary potentials, is $n \pi$, with $n$ the number of bound states. This induces a tendency for the phase shifts to start decreasing from threshold. This tendency has been used in lattice calculations \cite{newsasaki} with the help of L\"uscher's theorem to induce that one has indeed bound states in the infinite volume case. Note, however, that as found here, this feature alone would not tell us wheather the bound state is dynamically generated from $KD$ or a genuine state. Furthermore, we have checked explicitly that a potential of the CDD type does not fulfill the Levinson theorem., since we find that the phase shift decreases moderately as we increase the energy from threshold but then it starts increasing and hence does not decrease to $\delta(0)-\pi$ (see Fig.~\ref{fig:shiftscddfit})

\subsection{Case of a composite state analyzed in terms of a CDD potential}
\begin{figure}
\includegraphics[width=0.65\textwidth]{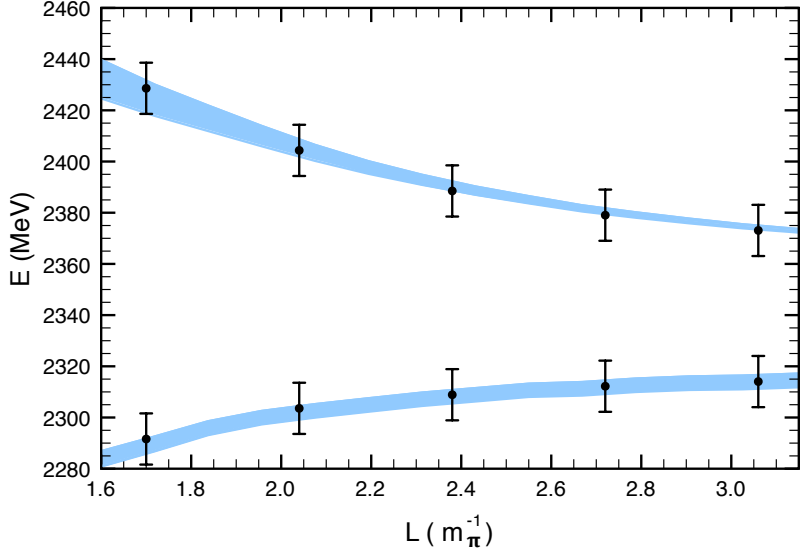}
\caption{Fits to the ``data" of Fig. \ref{fig:levelstwo} using a potential of the type of Eq.~(\ref{cdd}) that contains a CDD pole.}
\label{fig:fitcddtwochannel}
\end{figure}

\begin{figure}
\includegraphics[width=0.65\textwidth]{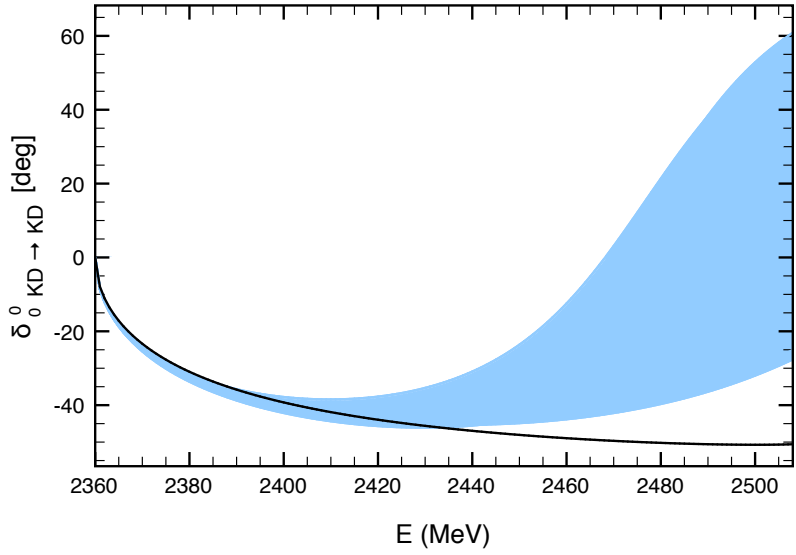}
\caption{Phase shifts from the fits with a CDD pole to the first two levels obtained with the coupled channels approach of Ref.~\cite{daniel} shown in Fig. \ref{fig:fitcddtwochannel}. The solid line corresponds to the results with the two channel analysis of Fig.~\ref{fig:phaseshiftstwo}.}
\label{fig:shiftcddtwochannel}
\end{figure}

It is also interesting to see if we could get a fit to the levels in the box obtained with the coupled channels approach of Ref.~\cite{daniel}, shown in Fig. \ref{fig:levelstwo}, by using the potential of Eq.~(\ref{cdd}) that contains a CDD pole. Considering these energy levels as ``lattice data"  we fit them with the potential of Eq.~(\ref{cdd}) and find the results shown in Fig. \ref{fig:fitcddtwochannel}. As we can see, the fit to the data is quite good.
It is also interesting to plot the $KD$ phase shifts obtained with this potential. In Fig. \ref{fig:shiftcddtwochannel} we show the results for the phase shifts obtained with the CDD potential. We can see that the phase shifts obtained up to 2400 MeV coincide with those obtained with the two channel potential. After this, there is a large dispersion of the results and they divert significantly from those obtained with the two channel approach. Certainly, the use of an additional ``lattice" level would put big constraints in the phase shifts in that region helping us decide between the two options. 

However, it is interesting to analyse the results obtained: we find the CDD pole at the energy of about 2500 MeV, far away from the bound state of the $D_{s^*0}(2317)$ state.  If one restricts oneself to low energies, the fit with the CDD would be acceptable. Yet, the fact that the CDD pole has appeared so far away from the energy of the $D_{s^*0}(2317)$ state is telling us that the data do not want a CDD pole being responsible for this state. The CDD pole far away in this case simply generates a smooth energy dependent potential in the region of the low energies. The interesting thing is that if we calculate now $Z$ from  Eq.~(\ref{sumrulecdd}) we find that $Z\sim 0.15$, indicating that the bound state is basically a $KD$ bound state, with the precision that the limited low energy data provide. A more precise determination would require the consideration of an additional level of the box. 

\subsection{Case of a genuine state analyzed with an ordinary potential}
\begin{figure}[h!]
\includegraphics[width=0.65\textwidth]{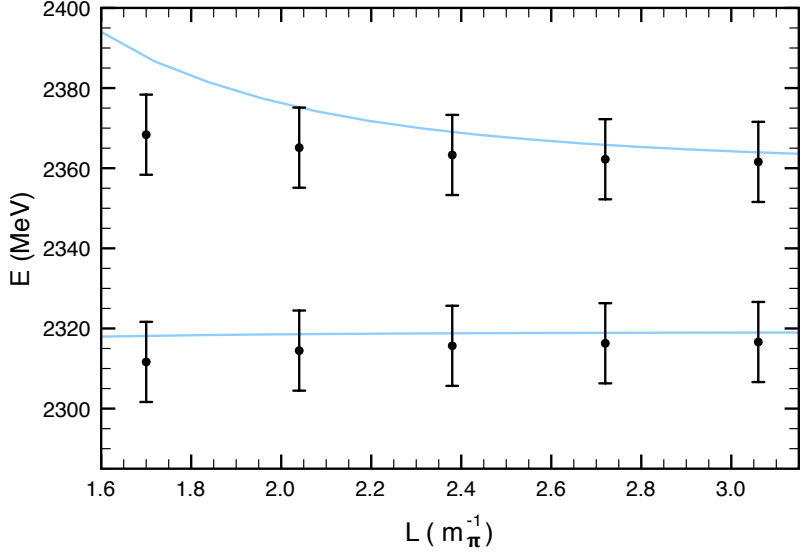}
\caption{Fit to the levels produced with a CDD potential using the potential of Eq.~(\ref{fitv}) with one channel.}
\label{fig:normalvtocdd}
\end{figure}

 It is also interesting to perform another test. Let us assume we had the lattice data which correspond to the levels generated with the CDD potential in Fig. \ref{fig:levelscdd} and we would like to fit them with the smooth potential of Eq.~(\ref{fitv}). The fit is bad, the $\chi^2$ is now of the order of 6 and the fit to the data can be seen in Fig. \ref{fig:normalvtocdd}. The quality of the fit is worse than in all the other fits that we have produced. Certainly, a better precision in the lattice data, and/or the addition of extra data at smaller box sizes would tell us that the fit is actually very bad. The addition of a third level would certainly help to disregard this solution. The exercise has also served to show the importance of the scattering data to determine the nature of a bound state. 
 
\begin{figure}[h!]
\includegraphics[width=0.65\textwidth]{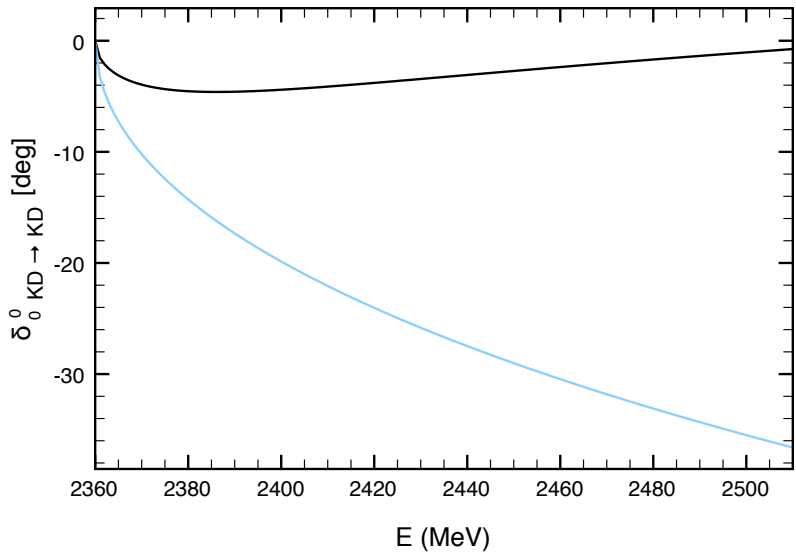}
\caption{Phase shifts derived from the fit with the potential of Eq.~(\ref{fitv}) with one channel to the levels produced with the CDD potential (lower curve). The upper line is the phase shift obtained with the CDD potential shown for reference.}
\label{fig:shiftcddwithnormal}
\end{figure}

 To finalize the discussion, we present in Fig. \ref{fig:shiftcddwithnormal} the results of the phase shifts that this fit would generate. As we can see, the results are very different from those generated from the CDD potential, such that if more precision is demanded to the lattice data such that they can produce the phase shifts with accuracy, this type of fit would be easily ruled out.   
  
\section{Conclusions}
In summary we have addressed three problems in the present work. The first one is to use the unitary coupled channel method in finite volume, generating the levels in a finite size box as a function of $L$. The second problem is the inverse problem: we assume that the results obtained before would correspond to results given by lattice calculations. From them we would like to obtain bound states and scattering states in the case of the infinite volume. We found that the two channel method proposed by us works well and is independent of the regularization scale used, which obviously we do not know when the lattice data are provided to us. If we stick to low energies of the $KD$ system we also see that the one channel analysis is quite good, but as we approach the threshold of $\eta D_s$, the two channel analysis provides better phase shifts.  The third issue faced in the work is whether we could say something from the lattice results concerning the nature of the bound state obtained. We found that indeed it was possible to induce some information on the nature of this state as basically a molecule of $KD$, by inspecting the spectra in finite volume. However we needed two levels to get an answer to this problem. We carried out a useful exercise  looking for the different alternatives in the analysis. We showed that if the bound state is a composite state of $KD$, the analysis using an ordinary potential and the Weinberg condition provided $Z\sim 0$ indicating the compositeness of the state. Alternatively we tried to analyze the lattice data in terms of a CDD potential that could accommodate a non molecular state and we could find a good fit to the data but with an artificial pole at an energy very far away, such that the potential was essentially constant in the region of interest and the $Z$ value for overlap with a genuine state was small, compatible with zero within the precision that the data allow. 

Similarly we played the exercise of assuming that the lattice data would correspond to a CDD potential that represents mostly a genuine state, with $Z\sim 0.9$. The analysis with a CDD potential provided a good fit to the data and returned this value of $Z$. On the contrary, attempts to fit these data with an ordinary potential failed, leading to unacceptable values of the $\chi^2$.

All these exercises show that the data provided by QCD lattice calculations on energy levels in a box contain the information to decide on the nature of the bound state $D_{s^*0}(2317)$. With the amount of data and precision suggested here one can obtain $Z$, the probability to have a genuine state, with a precision of about of about $\Delta Z\sim 0.1$. This is already quite a good precision at a time where the discussion goes on around the nature of some states at a qualitative level. Certainly more information and better precision on the lattice results could improve this error if desired.

\section*{Acknowledgments}
We would like to thank S. Sasaki for useful comments and M. D\"oring for a careful reading of the manuscript and useful suggestions. This work is partly supported by DGICYT contracts  FIS2006-03438,
 the Generalitat Valenciana in the program Prometeo and 
the EU Integrated Infrastructure Initiative Hadron Physics
Project under Grant Agreement n.227431 and by the Grants-in-Aid for Scientific Research (No. 22740161 and No. 22105507). This work was done in part under the Yukawa International Program for Quark-hadron Sciences (YIPQS).The work of A. M. T. is supported by the
Grant-in-Aid for the Global COE Program ``The Next Generation of Physics, Spun from Universality
and Emergence" from the Ministry of Education, Culture, Sports, Science and Technology
(MEXT) of Japan. This work is partly supported by The National Natural Science Foundation of
China (No.10975068) and the Scientific Research Foundation of Liaoning Education Department (No.2009T055).


\begin{thebibliography}{99}

\bibitem{Nakahara:1999vy}
  Y.~Nakahara, M.~Asakawa, T.~Hatsuda,
  Phys.\ Rev.\  {\bf D60} (1999)  091503;

  K.~Sasaki, S.~Sasaki and T.~Hatsuda,
  Phys.\ Lett.\  B {\bf 623} (2005) 208.
  
\bibitem{Mathur:2006bs}
  N.~Mathur, A.~Alexandru, Y.~Chen {\it et al.},
  Phys.\ Rev.\  {\bf D76} (2007) 114505.
 
\bibitem{Basak:2007kj}
  S.~Basak, R.~G.~Edwards, G.~T.~Fleming {\it et al.},
  Phys.\ Rev.\  {\bf D76} (2007) 074504.

\bibitem{Bulava:2010yg}
  J.~Bulava, R.~G.~Edwards, E.~Engelson {\it et al.},
  Phys.\ Rev.\  {\bf D82} (2010) 014507.
  
\bibitem{Morningstar:2010ae}
  C.~Morningstar, A.~Bell, J.~Bulava {\it et al.},
  AIP Conf.\ Proc.\  {\bf 1257} (2010) 779.

 
\bibitem{Foley:2010te}
  J.~Foley, J.~Bulava, K.~J.~Juge {\it et al.},
  AIP Conf.\ Proc.\  {\bf 1257} (2010) 789.
  

\bibitem{Alford:2000mm}
  M.~G.~Alford and R.~L.~Jaffe,
  Nucl.\ Phys.\  B {\bf 578} (2000) 367.


\bibitem{Kunihiro:2003yj}
  T.~Kunihiro, S.~Muroya, A.~Nakamura, C.~Nonaka, M.~Sekiguchi and H.~Wada
                  [SCALAR Collaboration],
  Phys.\ Rev.\  D {\bf 70} (2004) 034504.

\bibitem{Suganuma:2005ds}
  F.~Okiharu {\it et al.},
  arXiv:hep-ph/0507187;


  H.~Suganuma, K.~Tsumura, N.~Ishii and F.~Okiharu,
  PoS {\bf LAT2005} (2006) 070;
  Prog.\ Theor.\ Phys.\ Suppl.\  {\bf 168} (2007) 168.

\bibitem{Hart:2006ps}  
  C.~McNeile and C.~Michael  [UKQCD Collaboration],
  Phys.\ Rev.\  D {\bf 74} (2006) 014508;


  A.~Hart, C.~McNeile, C.~Michael and J.~Pickavance  [UKQCD Collaboration],
  Phys.\ Rev.\  D {\bf 74} (2006) 114504.

\bibitem{Wada:2007cp}
  H.~Wada, T.~Kunihiro, S.~Muroya, A.~Nakamura, C.~Nonaka and M.~Sekiguchi,
  Phys.\ Lett.\  B {\bf 652} (2007) 250.
  
\bibitem{Prelovsek:2010gm}  
S.~Prelovsek, C.~Dawson, T.~Izubuchi, K.~Orginos and A.~Soni,
  Phys.\ Rev.\  D {\bf 70} (2004) 094503;


  S.~Prelovsek, T.~Draper, C.~B.~Lang, M.~Limmer, K.~F.~Liu, N.~Mathur 
  and D.~Mohler,
  arXiv:1002.0193 [hep-ph];
  arXiv:1005.0948 [hep-lat].

\bibitem{luscher}
  M.~L\"uscher,
  Commun.\ Math.\ Phys.\  {\bf 105} (1986) 153 (1986).

\bibitem{Luscher:1990ux}
  M.~L\"uscher,
  Nucl.\ Phys.\  B {\bf 354} (1991) 531.

 
\bibitem{Liu:2005kr}
  C.~Liu, X.~Feng and S.~He,
  Int.\ J.\ Mod.\ Phys.\  A {\bf 21} (2006) 847.

\bibitem{Lage:2009zv}
  M.~Lage, U.-G.~Mei{\ss}ner and A.~Rusetsky,
  Phys.\ Lett.\  B {\bf 681} (2009) 439.
  
\bibitem{akaki}
  V.~Bernard, M.~Lage, U.-G.~Mei{\ss}ner and A.~Rusetsky,
  JHEP {\bf 1101} (2011) 019.
  
\bibitem{Bernard:2008ax}
  V.~Bernard, M.~Lage, U.-G.~Mei{\ss}ner and A.~Rusetsky,
  JHEP {\bf 0808} (2008) 024.
  
\bibitem{misha}
  M.~Doring, U.~-G.~Meissner, E.~Oset, A.~Rusetsky,
  Eur.\ Phys.\ J.\ A {\bf 47}, 139 (2011).
  
  
\bibitem{a}
  G.~Janssen, B.~C.~Pearce, K.~Holinde and J.~Speth,
  Phys.\ Rev.\  D {\bf 52}, 2690 (1995).
  
\bibitem{b}
 M.~Doring, J.~Haidenbauer, U.~-G.~Meissner, A.~Rusetsky,
Eur. Phys. J in print, arXiv:1108.0676 [hep-lat].
  
  \bibitem{Michaelnew}
  M.~Doring and U.~G.~Meissner,
  arXiv:1111.0616 [hep-lat].
  
\bibitem{Kolomeitsev:2003ac}
  E.~E.~Kolomeitsev, M.~F.~M.~Lutz,
  Phys.\ Lett.\  {\bf B582}, 39-48 (2004).
  
  
\bibitem{Hofmann:2003je}
  J.~Hofmann, M.~F.~M.~Lutz,
  Nucl.\ Phys.\  {\bf A733}, 142-152 (2004).
  

\bibitem{Guo:2006fu}
  F.~-K.~Guo, P.~-N.~Shen, H.~-C.~Chiang, R.~-G.~Ping, B.~-S.~Zou,
  Phys.\ Lett.\  {\bf B641}, 278-285 (2006).


\bibitem{daniel}
  D.~Gamermann, E.~Oset, D.~Strottman, M.~J.~Vicente Vacas,
  Phys.\ Rev.\  {\bf D76}, 074016 (2007).

 \bibitem{juan}
J.~M.~Flynn, J.~Nieves,
Phys.\ Rev.\  {\bf D75}, 074024 (2007).

\bibitem{hanhart2}  
 F.~-K.~Guo, C.~Hanhart, U.~-G.~Meissner,
 Eur.\ Phys.\ J.\  {\bf A40 } (2009)  171-179.
  
  \bibitem{guo1}
  F.~-K.~Guo, C.~Hanhart, S.~Krewald, U.~-G.~Meissner,
  Phys.\ Lett.\  {\bf B666}, 251-255 (2008).

  
\bibitem{weinberg}
  S.~Weinberg,
  Phys.\ Rev.\  {\bf 137}, B672-B678 (1965).
  
\bibitem{hanhart}
  V.~Baru, J.~Haidenbauer, C.~Hanhart, Y.~.Kalashnikova, A.~E.~Kudryavtsev,
  Phys.\ Lett.\  {\bf B586}, 53-61 (2004).
 
  

  

\bibitem{ssasaki} S. Sasaki, private communication.

\bibitem{ollerulf}
  J.~A.~Oller, U.~G.~Meissner,
  Phys.\ Lett.\  {\bf B500}, 263-272 (2001).
 

\bibitem{bennhold}
  E.~Oset, A.~Ramos, C.~Bennhold,
  Phys.\ Lett.\  {\bf B527}, 99-105 (2002).
 
  
  
\bibitem{npa}
  J.~A.~Oller, E.~Oset,
  Nucl.\ Phys.\  {\bf A620}, 438-456 (1997).
  

\bibitem{c}
For tabulated numbers and further references see, e.g., The On-Line Encyclopedia of Integer Sequences, http://oeis.org/A005875.

\bibitem{d}
J.~A.~Oller, E.~Oset and J.~R.~Pelaez,
Phys.\ Rev.\  D {\bf 59} (1999) 074001
[Erratum-ibid.\  D {\bf 60} (1999) 099906]
[Erratum-ibid.\  D {\bf 75} (2007) 099903].

\bibitem{Ben}
P. R. Bevinton, Data reduction and error analysis for the physical science, McGraw-Hill, 1969.
\bibitem{e}
  T.~Hyodo, D.~Jido, A.~Hosaka,
  Phys.\ Rev.\  {\bf C78}, 025203 (2008).

\bibitem{friman}
A.~Schwenk, B.~Friman and G.~E.~Brown,
  Nucl.\ Phys.\ A {\bf 713}, 191 (2003).
\bibitem{hyodo}
  T.~Hyodo, W.~Weise,
  Phys.\ Rev.\  {\bf C77}, 035204 (2008).
 
 
 \bibitem{guo2}
  M.~Cleven, F.~-K.~Guo, C.~Hanhart, U.~-G.~Meissner,
  Eur.\ Phys.\ J.\  {\bf A47}, 19 (2011).
 

\bibitem{arriola}
  D.~Gamermann, J.~Nieves, E.~Oset, E.~Ruiz Arriola,
  Phys.\ Rev.\  {\bf D81}, 014029 (2010).
   
\bibitem{jidohyodo}
  T.~Sekihara, T.~Hyodo, D.~Jido,
  Phys.\ Rev.\  {\bf C83}, 055202 (2011).
  
 \bibitem{two}
  T.~Hyodo, D.~Jido, A.~Hosaka,
  Phys. Rev. C in print, arXiv:1108.5524 [nucl-th].
\bibitem{leonardo}
  L.~Castillejo, R.~H.~Dalitz, F.~J.~Dyson,
  Phys.\ Rev.\  {\bf 101}, 453-458 (1956).
  

\bibitem{newsasaki}
  S.~Sasaki, T.~Yamazaki,
  Phys.\ Rev.\  {\bf D74}, 114507 (2006).
 

\bibitem{sasakifriend}
  T.~Yamazaki, Y.~Kuramashi, A.~Ukawa, f.~t.~P.~-C.~Collaboration,
  arXiv:1105.1418 [hep-lat].







\end{thebibliography}
\end{document}